\documentstyle[12pt]{article}

\newcommand{\nc}{\newcommand}
\newcounter{romzahl}
\nc{\Rz}[1]{\setcounter{romzahl}{#1} \Roman{romzahl}}

\nc{\para}{{\scriptscriptstyle\parallel}}
\nc{\senk}{{\scriptscriptstyle\perp}}

\nc{\be}{\begin{eqnarray}}
\nc{\ee}{\end{eqnarray}}
\nc{\nn}{\nonumber}
\nc{\ts}{\textstyle}

\nc{\mhspace}[1]{{\mbox{\hspace{#1}}}}
\nc{\mvspace}[1]{{\mbox{\vspace{#1}}}}
\nc{\mscript}[1]{_{\mbox{\scriptsize \it #1}}}

\nc{\ec}[3]{{C_{#1_#3}^{#2^#3_1#2^#3_2}}}
\nc{\er}[3]{R^{\, {#1_#3}}_{#2^#3_1#2^#3_2}}
\nc{\ecklamauf}[3]{C_{\,{#1_#3}}^{\{{#2^#3_1#2^#3_2}}}
\nc{\erklamauf}[3]{R^{#1_#3}_{\{{#2^#3_1#2^#3_2}}}
\nc{\ecklamzu}[3]{C_{#1_#3}^{#2^#3_1#2^#3_2\}_{as}}}
\nc{\erklamzu}[3]{R^{#1_#3}_{#2^#3_1#2^#3_2\}_{as}}}
\nc{\zc}[3]{\ec{#1}{#2}{1} \ldots \ec{#1}{#2}{#3}}
\nc{\zr}[3]{\er{#1}{#2}{1} \ldots \er{#1}{#2}{#3}}
\nc{\zcas}[3]{\ecklamauf{#1}{#2}{1} \ldots \ecklamzu{#1}{#2}{#3}}
\nc{\zras}[3]{\erklamauf{#1}{#2}{1} \ldots \erklamzu{#1}{#2}{#3}}
\nc{\gen}[1]{\mbox{\bf\sf #1}}
\nc{\funkop}[1]{{\cal #1}}
\nc{\frak}[1]{{\cal #1}}
\nc{\dalembert}{{\mbox{\large$\Box$}}}
\nc{\hatdalembert}{{\mbox{\large$\hat{\Box}$}}}
\nc{\christoffel}[2]{\left\{\!\!\begin{array}{c}{#1}\\
   {#2}\end{array}\!\!\right\}}

\nc{\ket}[1]{\left|\,{#1}\,\right>}
\nc{\bra}[1]{\left<\,{#1}\,\right|}
\nc{\braket}[2]{\left<\,{#1}\mid{#2}\,\right>}
\nc{\varket}[1]{\left|\,{#1}\,\right)}
\nc{\varbra}[1]{\left(\,{#1}\,\right|}
\nc{\varbraket}[2]{\left(\,{#1}\mid{#2}\,\right)}

\nc{\lek}{\left[}
\nc{\rek}{\right]}
\nc{\lrk}{\left(}
\nc{\rrk}{\right)}
\nc{\lgk}{\left\{}
\nc{\rgk}{\right\}}
\nc{\luk}{\left.}
\nc{\ruk}{\right.}
\nc{\klr}[1]{\left(\,#1\,\right)}
\nc{\klg}[1]{\left\{\,#1\,\right\}}
\nc{\kle}[1]{\left[\,#1\,\right]}
\nc{\dokle}[1]{\kle{{\mhspace{-6.5pt}}\kle{{#1}}{\mhspace{-6.5pt}}}}

\nc{\ehoch}[1]{\exp \klg{#1}}
\nc{\dvier}{d^{^{\scriptstyle 4}} \!\!\! }
\nc{\ddrei}{d^{^{\scriptstyle \, 3}} \!\!\! }
\nc{\deltavier}{\delta^{^{\scriptstyle 4}} \!\! }

\nc{\SS}{{\cal S}}
\nc{\TT}{{\cal T}}
\nc{\HH}{{\cal H}}

\nc{\Sp}{S\!p\,}
\nc{\betr}[1]{\left\vert #1 \right\vert}
\nc{\norm}[1]{\left\Vert #1 \right\Vert}
\nc{\quer}[1]{\overline{#1}}
\nc{\matrixzz}[4]{\klr{\begin{array}{cc} #1 & #2 \\ #3 & #4
                  \end{array}}}
\nc{\ua}{\uparrow}
\nc{\da}{\downarrow}

\nc{\foda}[1]{\pi_{{\cal F}} \lrk \, #1 \, \rrk }
\nc{\dufoda}[1]{\pi_{{\cal F}^\ast } \lrk \, #1 \, \rrk }
\nc{\fovak}{\Omega _{{\cal F}}}
\nc{\dufovak}{\Omega _{{\cal F}^\ast }}
\nc{\fozu}[1]{\omega_{{\cal F}} \lrk \, #1 \, \rrk }
\nc{\dufozu}[1]{\omega_{{\cal F}^\ast } \lrk \, #1 \, \rrk }
\nc{\nfozu}{\omega_{{\cal F}}}
\nc{\ndufozu}{\omega_{{\cal F}^\ast}}
\nc{\omvw}[1]{\omega_ {v_{#1}w_{#1}}}

\nc{\hred}{H_{\mbox{\scriptsize red}}}
\nc{\heff}{H_{\mbox{\scriptsize eff}}}
\nc{\psibcs}{\psi_{_{BCS}}}

\nc{\ep}{\varepsilon}
\nc{\lam}{\lambda}
\nc{\sig}{\sigma}
\nc{\Lam}{\Lambda}
\nc{\om}{\omega}
\nc{\Om}{\Omega}
\nc{\al}{\alpha}
\nc{\ga}{\gamma}
\nc{\ka}{\kappa}
\nc{\vp}{\varphi}

\nc{\1}{{{\hbox{1}}\!{\hbox{l}}}}
\nc{\N}{{{\hbox{I}}\!{\hbox{N}}}}
\nc{\R}{{{\hbox{I}}\!{\hbox{R}}}}
\nc{\Z}{{{\hbox{Z}\!\!\hbox{Z}}}}
\nc{\C}{\hbox{\rlap{$\,\,$\hbox{\vrule height1.5ex width0.12ex
          depth0ex}}C}}
\nc{\Q}{{\hbox{\rlap{$\,\,$%
          \hbox{\vrule height6pt width1pt depth0.1pt}}Q}}}

\nc{\avonc}{{\cal A}\lrk \C \rrk }
\nc{\avoncn}{{\cal A}\lrk \C^n \rrk }
\nc{\aueinsvonc}{{{\cal A}_{U\lrk 1\rrk }} \lrk \C \rrk }
\nc{\avonczwei}{{\cal A}\lrk \C^2 \rrk }
\nc{\auzwei}{{{\cal A}_{U\lrk 2\rrk }} \lrk \C^2 \rrk }

\nc{\ct}{\tilde{c}}

\nc{\dm}{d\mu}
\nc{\dmf}{d\mu_{_{{\cal F}}}}
\nc{\dmfast}{d\mu_{_{{\cal F}^\ast}}}
\nc{\dxi}{d\xi}
\nc{\dxiast}{d\xi^{\ast}}
\nc{\dxixi}{d\xi^{\ast}d\xi}
\nc{\dxil}{d\xi^{}_l}
\nc{\dxilast}{d\xi_l^{\ast}}
\nc{\dxir}{d\xi^{}_r}
\nc{\dxirast}{d\xi_r^{\ast}}
\nc{\dxixil}{d\xi_l^{\ast}d\xi^{}_l}
\nc{\dxixir}{d\xi_r^{\ast}d\xi^{}_r}
\nc{\dxixilr}{d\xi_r^{\ast}d\xi^{}_rd\xi_l^{\ast}d\xi^{}_l}
\nc{\xiast}{\xi^{\ast}}
\nc{\xil}{\xi_l^{}}
\nc{\xilast}{\xi_l^{\ast}}
\nc{\xir}{\xi_r^{}}
\nc{\xirast}{\xi_r^{\ast}}
\nc{\klxi}{\klr{\xi,\,\xi^{\ast}}}
\nc{\klxilr}{\klr{\xil,\,\xilast,\,\xir,\,\xirast}}

\nc{\etal}{\eta_l^{}}
\nc{\etalast}{\eta_l^{\ast}}
\nc{\etar}{\eta_r^{}}
\nc{\etarast}{\eta_r^{\ast}}
\nc{\kletalr}{\klr{\etal,\,\etalast,\,\etar,\,\etarast}}

\nc{\cc}[3]{C^{{#1}{#2}}_{#3}}
\nc{\rr}[3]{R_{{#1}{#2}}^{\: #3}}
\nc{\vj}[2]{j_{I_1^{#1}}j_{I_2^{#1}}\ldots j_{I_1^{#2}}j_{I_2^{#2}}}
\nc{\abl}[2]{\frac{\delta }{\delta {#1}_{#2}}}
\nc{\jabl}[1]{\frac{\delta }{\delta j_{I_{#1}}}}
\nc{\habl}[1]{\frac{\delta }{\delta h_{K_{#1}}}}
\nc{\zjabl}[1]{\frac{\delta }{\delta j_{I^{#1}_2}} \frac{\delta
}{\delta j_{I^{#1}_1}}}
\nc{\babl}[1]{\frac{\delta }{\delta b_{k_{#1}}}}
\nc{\bablk}[1]{\frac{\delta }{\delta b_{k{#1}}}}
\nc{\babll}[1]{\frac{\delta }{\delta b_{l{#1}}}}
\nc{\pabl}[1]{\frac{\partial }{\partial {#1}}}
\nc{\aket}[1]{\ket{{\cal A}\; [\:{#1}\:]}}
\nc{\bket}[1]{\ket{{\cal B}\; [\:{#1}\:]}}
\nc{\bbra}[1]{\bra{{\cal B}\; [\:{#1}\:]}}
\nc{\gbra}[1]{\bra{{\cal G}\; [\:{#1}\:]}}
\nc{\atildeket}[1]{\ket{\tilde{{\cal A}}\; [\:{#1}\:]}}
\nc{\tildefket}[1]{\ket{\tilde{{\cal F}}\; [\:{#1}\:]}}
\nc{\fket}[1]{\ket{{\cal F}\; [\:{#1}\:]}}
\nc{\tket}[1]{\ket{{\cal T}\; [\:{#1}\:]}}
\nc{\s}[2]{S^{#1_1\ldots #1_n}_{#2_1\ldots #2_n}}
\nc{\skl}{S^{k_1 \ldots k_n}_{l_1 \ldots \, l_n}}
\nc{\gammas}{\Gamma^{\, l \, l'}_{kk'} \, S^{m_1 \ldots m_n \: k
            \quad \: k'}_{m'_1 \ldots m'_n m'_{n+1} m'_{n+2}}}

\begin{document}
\thispagestyle{empty}
\vspace*{2cm}
\centerline{\Large \bf On an inconsistency in}
\vspace{0.5\baselineskip}
\centerline{\Large \bf path integral bosonization}
\vspace{2cm}
\begin{center}
R.~Kerschner\\
Institut f\"ur Theoretische Physik, University of T\"ubingen\\
Auf der Morgenstelle 14, 72076 T\"ubingen, Germany
\end{center}
\vspace{2cm}
\begin{center}
\parbox{14cm}{\begin{tabbing}
00000000\=00000000000\=00000000000000000000000000000000000000000000000\kill
                  \>            \>Formalism\\
                  \>  03.65.Ge  \>Quantum theory; quantum mechanics: \\
                  \>            \>Solutions of wave equations: bound states\\
                  \>  03.70.+k  \>Theory of quantized fields\\
                  \>  11.15.--q \>Gauge field theories\\
                  \>  12.38.--t \>Quantum chromodynamics\\
                  \>  14.40.--n \>Mesons and meson resonances
                    \end{tabbing}}
\end{center}

\newpage
\thispagestyle{empty}
\vspace*{10cm}
\centerline{\Large \bf Abstract}
\vspace{2\baselineskip}
\centerline{\parbox{10cm}{
A critically discerning discussion of path integral bosonization
is given. Successively evaluating the conventional path integral
bosonization of QCD it is shown without any approximations that gluons
must be composed of two quarks. This contradicts the
fundamentals of QCD, where quarks and gluons are independent
fields. Furthermore, bosonizing the
Fierz reordered effective four quark interaction term yields
gluons, too. Colorless ``mesons'' are shown to be Fierz
equivalent to a submanifold of gluons. The results obtained are
not specific to QCD, but apply to other models as well.}}
\setcounter{page}{0}

\newpage
\oddsidemargin1.7cm
\section{Introduction}
Due to constrained dynamical variables the canonical
quantization of gauge theories is rather involved. It is
commonly believed that a (renormalizable) quantum field theory
can equivalently be expressed by a path integral. Restricting
ourselves to quantum chromodynamics (QCD), the path integral is
formulated in terms of unobservable fields, i.~e.~quarks
and gluons. The observable quantities are considered to be
colorless composite particles which are composed of two or three
quarks. Thus the path integral quantization of QCD is only
complete if it is supplied by a theory of formation and dynamics
of composite particles. In the past several efforts were made to
develop such effective composite particle theories, e.g.
\cite{pibos}, \cite{rei}.
\nocite{kleinert76}
\nocite{eguchisugawara74} \nocite{kikkawa76}
\nocite{reinhardt91} \nocite{ebertreinhardt86a}
\nocite{cahillroberts85} \nocite{prc87} \nocite{aann87}
\nocite{furlanraczka83}
The general strategy
is to eliminate the gluons in favour of an effective quark
interaction theory, which is then bosonized in order to obtain
phenomenological meson theories. The problem of an exact
treatment of the gluon selfinteraction has recently been solved
\cite{rei}
\nocite{reinhardt90a} \nocite{reinhardt91} using the field strength
approach \cite{halp} \nocite{halpern77a} to QCD.
To derive our results we shall make use of this
formalism since it admits exact calculations.

Many critical remarks on path integrals have already been given by
e.~g.~Rivers \cite{rivers}. In addition to these objections it
will be shown that no serious composite
particle theory can be obtained from path integral bosonization.

\section{Gluons as two-quark composites}
As our starting point we consider the Euclidean path integral
\be
Z = \int D\kle{\bar c} D\kle{c} D\kle{\bar q} D\kle{q} D\kle{A}
\ehoch{-S_{\mbox{\scriptsize \it QCD}} -
S_{\mbox{\footnotesize \it gf}} } \; ,
\label{gl1}
\ee
where
\be
S_{\mbox{\scriptsize \it QCD}}
& = & \int \dvier x \; \klg{\bar q \klr{\ga_\mu \partial_\mu + m} q
+ \frac{1}{4} F^a_{\mu\nu} F^a_{\mu\nu} - ig \bar q \ga_\mu t^a
q A^a_\mu }
\nn \\
F^a_{\mu\nu} & = & \partial_\mu A^a_\nu - \partial_\nu A^a_\mu +
g f_{abc} A^b_\mu A^c_\nu
\label{gl2}
\ee
is the action and field strength tensor of QCD and we have
chosen a covariant gauge fixing
\be
S_{\mbox{\footnotesize \it gf}} = \int \dvier x \; \klg{
-\frac{1}{2\xi} \klr{\partial_\mu A^a_\mu}^2 + i \bar c_a
\klr{\Box \delta_{ab} - g f_{abc} \partial_\mu A^c_\mu} c_b }
\label{gl3}
\ee
with $\bar c$ and $c$ being the Faddeev--Popov ghost
fields, cf.~\cite{chengli}.

Multiplying equation (\ref{gl1}) with
\be
1 & = & \int D\kle{G} D\kle{\psi} \delta \klr{G^a_{\mu\nu} -
F^a_{\mu\nu} (A) } \delta \klr{\psi^a - \partial_\mu A^a_\mu }
\nn\\
& = & \int D\kle{G} D\kle{T} D\kle{\psi} D\kle{\phi} \exp \bigg\{
\frac{i}{2} \int \dvier x \; T^a_{\mu\nu} \klr{G^a_{\mu\nu} -
F^a_{\mu\nu} (A) } +
\nn\\
&& \mhspace{3cm} +i \int \dvier x \; \phi^a \klr{\psi^a -
\partial_\mu A^a_\mu}\bigg\}
\label{gl4}
\ee
allows us to introduce the field strength tensor as an
independent variable. The $\psi$--field only serves to simplify
(\ref{gl3}) and the following calculation. Due to the
$\delta$--distributions we can replace $F^a_{\mu\nu} (A)$ by
$G^a_{\mu\nu}$ and $\partial_\mu A^a_\mu$ by $\psi^a$ in
(\ref{gl1}). After integrating out the $G$--field we obtain
\be
Z = \int D\kle{\bar q} D\kle{q} D\kle{T} D\kle{\phi} Z_1
\kle{\bar q, q, T, \phi} Z_2 \kle{\bar q, q, T, \phi}
\label{gl5}
\ee
with
\be
Z_1 & = & \int D\kle{\bar c} D\kle{c} D\kle{\psi} \exp \bigg\{- \int
\dvier x \; \Big[\bar q \klr{\ga_\mu \partial_\mu + m} q +
\frac{1}{4} T^a_{\mu\nu} T^a_{\mu\nu} +
\nn\\
&& \mhspace{2cm} - \frac{1}{2\xi} \psi^a \psi^a
- i \phi^a \psi^a +i \bar c_a \klr{\Box
\delta_{ab} - g f_{abc} \psi^c} c_b\Big]\bigg\}
\label{gl6a} \\
Z_2 & = & \int D\kle{A} \exp \bigg\{- \int \dvier x \; \Big[-ig \bar q
\ga_\mu t^a q A^a_\mu + i \phi^a \partial_\mu A^a_\mu + i
T^a_{\mu\nu} \partial_\mu A^a_\nu +
\nn\\
&& \mhspace{3cm} + \frac{i}{2} g T^a_{\mu\nu}
f_{abc} A^b_\mu A^c_\nu\Big]\bigg\} \; .
\label{gl6b}
\ee
For the further evaluation we can concentrate on (\ref{gl6b}).
Performing partial integrations with respect to the space-time
coordinates and integrating out
the $A$--fields gives
\be
Z_2 = \klr{\det \tilde{T}}^{-\frac{1}{2}} \ehoch{\frac{1}{2}
\int \dvier x \; j^a_\mu (\tilde{T}^{-1})^{ab}_{\mu\nu}
j^b_\nu} \; ,
\label{gl8}
\ee
with $j^a_\mu$ and $\tilde{T}^{ab}_{\mu\nu}$ being defined by
\be
j^a_\mu & = & i g \bar q \ga_\mu t^a q + i R^a_\mu \klr{\phi, T}
\nn\\
R^a_\mu \klr{\phi, T} & = & \partial_\mu \phi^a + \partial_\nu
T^a_{\nu\mu}
\nn\\
\tilde{T}^{bc}_{\mu\nu} & = & i g T^a_{\mu\nu} f_{abc} \; .
\label{gl9}
\ee
The effective action in (\ref{gl8}) contains a four quark
interaction term, coupled to the (inverse) field strength
tensor. It is generally argued, that this interaction can be
replaced by a contact force at low energy, yielding a
Nambu--Jona-Lasinio type of model. These models have
extensively been studied for the purpose of deriving effective
mesonic actions in the low energy regime, e.g. \cite{njl}.
\nocite{nambujonalasinio61a}
\nocite{klevansky92}
Hence hopefully (\ref{gl8}) and (\ref{gl5})
will give a microscopic derivation of effective meson theories
from QCD, with the mesons being $\bar qq$--composites.
Depending on the energy scale we expect to observe form factors,
exchange force corrections stemming from the quark level,
and excited meson states.

Obviously there are several (equivalent) possibilities to
select quark pairings for the bosonization of
(\ref{gl8}). The most elucidatory way is
to directly choose $g\bar q \ga_\mu t^a q \rightarrow
\xi^a_\mu$, although the composites are no color singlets in
that case. Since one knows that observable mesons must be color
singlets this choice is not standard. Rather one usually applies
a Fierz reordering to (\ref{gl8}) before bosonizing it. However,
the Fierz transformation is a strict identity on the quark level
and we will show that it possesses an exact counterpart on the
bosonized level. On account of their coloredness we shall call
the above composites pseudo-mesons and postpone the discussion of the
effects of a Fierz reordering to the following section.

To simplify notation we combine Lorentz, color, and flavor
degrees of freedom to a single index
\be
\klr{\ga_\mu}_{l_1l_2} \klr{t^a}_{c_1c_2} \klr{\1}_{f_1f_2} =
\Lambda^A_{i_1i_2} \; .
\label{gl10}
\ee
Equation (\ref{gl8}) is then bosonized\footnote{This bosonization
technique is fully equivalent to the one used by other authors
\cite{pibos}, where one multiplies the path integral with a
Gauss integral and subsequently applies a shift in order to
eliminate the four quark interaction term.} by multiplying it
with
\be
1 & = & \int D \kle{\xi} \delta \klr{ \xi^A- g \bar{q} \Lambda^A q}
\nn\\
& = & \int D \kle{\xi}\, D\kle{\eta} \ehoch{\int \dvier x \, i
\eta^A \klr{\xi^A- g\bar{q} \Lambda^A q}}
\label{gl13}
\ee
and replacing $g\, \bar{q} \Lambda^A q$ by $\xi^A$, which gives
\be
Z_2 & = & \int D \kle{\xi} D \kle{\eta} \klr{\det \tilde{T}}^{-
\frac{1}{2}} \exp \bigg\{ \int \dvier x \, \Big[ i
\eta^A \klr{\xi^A - g \bar{q} \Lambda^A q} +
\nn\\
&&\mhspace{2cm} - \frac{1}{2} \klr{\xi^A + R^A}
(\tilde{T}^{-1})^{AB}  \klr{\xi^B + R^B} \Big]
\bigg\} \; .
\label{gl14}
\ee
Herein the $\eta$--coordinates describe the pseudo-mesons, while
the $\xi$--coordinates have to be eliminated. Integrating out
the $\xi$--coordinates and making use of (\ref{gl10}) and
(\ref{gl9}) yields
\be
Z_2 & = & \int D\kle{\eta} \exp \bigg\{ \int \dvier x \; \Big[ -
\frac{i}{2} g T^a_{\mu\nu} f_{abc} \eta^b_\mu \eta^c_\nu +
\nn\\
&& \mhspace{1.5cm} +i \eta^a_\mu \partial_\mu \phi^a +i \eta^a_\mu
\partial_\nu T^a_{\nu\mu} -ig\bar q \ga_\mu t^a q
\eta^a_\mu \Big]  \bigg\} \; ,
\label{gl15}
\ee
since the determinant from the $\xi$--integration exactly
compensates the determinant from the $A$--integration.
Equation (\ref{gl15}) coincides
exactly with (\ref{gl6b}) after partial integration and
renaming of $\eta^a_\mu \rightarrow A^a_\mu$. Inserting
(\ref{gl15}) into (\ref{gl5}) we can integrate out $T$, $\phi$ and
$\psi$ to get (\ref{gl1}).

{\it Hence, the
pseudo-mesons are exactly the gluons (fixed in the same gauge).}
Thus, if composite particle actions are obtained from path integral
bosonization, then the gluons must be composite particles, too!
This exceeds by far pair creation and annihilation phenomena,
since there should be form factor effects and fermionic exchange
forces in gluon-gluon scattering. In addition, we should
have excited gluon states. As a matter of principle this
result contradicts the original assumptions of QCD, that gluons
are described as independent fields.

\section{The effect of a Fierz reordering}

Applying a Fierz transformation to the four quark interaction
term in (\ref{gl8}) yields (among others) a color singlet
channel of two-quark pairings. Bosonizing the resulting channels
will therefore contain bosonic coordinates with the quantum
numbers of phenomenological mesons. Hence one might believe
that these coordinates will describe mesons. Subsequently we
will show that this is not the case.

Extending the $\Lambda$--matrices to a complete and
orthonormalized set of matrices $M^r$
\be
tr \klr{M^r M^{r'}}  =  \delta_{rr'} \; , \qquad
\sum_r M^r_{i_1i_2} M^r_{i_3i_4} & = & \delta_{i_1i_4}
\delta_{i_2i_3}
\label{gl17}
\ee
the Fierz transformation of (\ref{gl8}) reads
\be
Z_2 & = & \klr{\det \tilde{T}}^{-\frac{1}{2}} \exp \bigg\{
-\frac{1}{2} g^2 \bar q M^r q (\tilde{T}^{-1})^{AB} m^{AB}_{rs}
\bar q M^s q +
\nn\\
&& \mhspace{2cm} - g R^A (\tilde{T}^{-1})^{AB} \bar q \Lambda^B
q -\frac{1}{2} R^A (\tilde{T}^{-1})^{AB} R^B \bigg\} \; ,
\label{gl18}
\ee
where the Fierz matrix is given by\footnote{One might also consider
other Fierz transformations, but none change the following line
of arguments.}
\be
m^{AB}_{rs} = -tr \klr{\Lambda^A M^r \Lambda^B M^s} \; .
\label{gl19}
\ee
Bosonizing (\ref{gl18}) by multiplication with
\be
1 & = & \int D \kle{\xi} \, \delta \klr{\xi^r - g\bar q M^r q}
\nn\\
& = & \int D \kle{\xi} \, D\kle{\eta} \ehoch{\int \dvier x \,
i\eta^r \klr{\xi^r-g\bar{q} M^r q}}
\label{gl19a}
\ee
yields
\be
Z_2 & = & \int D\kle{\xi} D\kle{\eta} \klr{\det
\tilde{T}}^{-\frac{1}{2}} \exp \bigg\{\int \dvier x \; \Big[ i
\eta^r \xi^r - \frac{1}{2} \xi^r
(\tilde{T}^{-1})^{AB} m^{AB}_{rs} \xi^s +
\nn\\
&& -ig \bar q M^r q \eta^r - R^A
(\tilde{T}^{-1})^{AB} \xi^B - \frac{1}{2} R^A
(\tilde{T}^{-1})^{AB} R^B \Big] \bigg\} \; ,
\label{gl20}
\ee
but contrary to (\ref{gl14}) the $\xi$--integration is now more
involved. This is due to the fact that the Fierz matrix
satisfies the projector equation
\be
m^{AB}_{rs} = m^{AB}_{r's'}\, P^{r's'}_{rs} \; ,
\label{gl21}
\ee
where the projector $P^{r's'}_{rs}$ is given by
\be
P^{r's'}_{rs} = m^{AB}_{rs}\, m^{AB}_{r's'} \; .
\label{gl22}
\ee
Therefore the kernel $(\tilde{T}^{-1})^{AB} m^{AB}_{rs}$ has
eigenvalues zero, effectively restricting the non-trivial domain
of the $\eta$--integration.

Nevertheless the $\xi$--integration can be performed
exactly by taking into consideration that the
$\xi$--coordinates are not ordinary boson coordinates, but
bosonized Grassmann variables, which must be integrated over by
corresponding Berezin rules. These rules are derived from
(\ref{gl19a}) and read\footnote{In order to prove this formula,
one must simply bosonize the right hand side of (\ref{neu1}) by
multiplication with (\ref{gl19a}).}
\be
\int D\kle{\xi} \, F \klr{\xi^r} = \int D\mu \kle{\bar{\al}, \al}
\, F \klr{\bar{\al} M^r \al} \; ,
\label{neu1}
\ee
where $\bar{\al}$ and $\al$ are anticommuting Grassmann
variables and the measure is
\be
D\mu \kle{\bar{\al}, \al} = D\kle{\bar{\al}} \, D \kle{\al} \,
\mu \klr{\bar{\al} M^r \al}
\ee
with
\be
\mu \klr{\bar{\al} M^r \al}  =  \kle{\int D\kle{\eta}\, \det
\klr{-i\eta^rM^r} \ehoch{\int \dvier x \,i \eta^r \bar{\al} M^r
\al} }^{-1} \;,
\ee
and $F$ denotes an arbitrary functional.
Hence the bosonization forces us to integrate out the
$\xi$--coordinates by the bosonized Berezin rules (\ref{neu1}).

Applying (\ref{neu1}) to equation (\ref{gl20}) yields
\be
Z_2 &=& \int D\kle{\xi}\, D\kle{\eta} \, \klr{\det
\tilde{T}}^{-\frac{1}{2}} \exp \bigg\{ \int  \dvier x \Big[
i\xi^r\eta^r -ig\bar{q} M^r q \eta^r +
\nn\\
&& - \frac{1}{2} \klr{\xi^A + R^A} (\tilde{T}^{-1})^{AB}
\klr{\xi^B + R^B} \Big]\bigg\} \; ,
\label{gl26}
\ee
if we make use of the bosonized Fierz identity
\be
\int  D\kle{\xi} F\klr{ {\ts -\frac{1}{2}} \xi^r
(\tilde{T}^{-1})^{AB} m^{AB}_{rs} \xi^s} = \int D\kle{\xi}
F \klr{ {\ts -\frac{1}{2}} \xi^A
(\tilde{T}^{-1})^{AB} \xi^B } .
\label{neu3}
\ee
This identity is a consequence of (\ref{neu1}) alone and allows
us to remove the eigenvalues zero in the kernel
$(\tilde{T}^{-1})^{AB} m^{AB}_{rs}$ from (\ref{gl20}). Since
(\ref{neu3}) is an identity, the zero eigenspace of the
projector (\ref{gl22}) is actually not in the domain of the
$\xi$--integration, if this integration is evaluated by the
bosonized Berezin rules (\ref{neu1}).

The determinant in (\ref{gl26}) can be expressed by
a boson path integral, which gives
\be
Z_2 &=& \int D\kle{A} \, D \kle{\xi} \, D\kle{\eta} \, \exp
\bigg\{ \int \dvier x \Big[ i\xi^r\eta^r - i \xi^B A^B +
\nn\\
&& -\frac{1}{2} \tilde{T}^{BC} A^B A^C -i R^B A^B -ig \bar{q}
M^r q \eta^r \Big]\bigg\} \; .
\label{gl25}
\ee
If the Fourier decomposition of the $\delta$--function in
(\ref{gl19a}) is to be consistent with exchanging the order of
$\xi$-- and $\eta$--integration, we must have
\be
\delta \klr{\chi^r} = \int D\kle{\xi} \ehoch{\int \dvier x \, i
\xi^r \chi^r} \;.
\ee
Integrating out the $\xi$--coordinates in (\ref{gl25}), we
therefore get
\be
Z_2 & = & \int D\kle{A} \, D\kle{\eta} \, \delta \klr{\eta^r -
\delta^r_B A^B } \,\times
\nn\\
&& \quad \times \exp
\bigg\{ \!\!-\frac{1}{2} \tilde{T}^{BC} A^B A^C -i R^B A^B -ig \bar{q}
M^r q \eta^r \Big]\! \bigg\}.
\label{neu2}
\ee
Either one of the $A$-- or $\eta$--integration yields
(\ref{gl15}) again.

Thus a Fierz reordering does not give any new results. Rather
the Fierz transformed integral (\ref{gl20}) conceals the
non-trivial domain of the $\eta$--integration, if the
$\xi$--coordinates are not integrated over by bosonized Berezin
rules.  Any ``approximation'' of solely selecting specific
channels and forgetting about the Grassmann properties of the
$\xi$--integration will definitely change the theory in a way
which is no longer consistent with QCD.

Taking the steps (\ref{neu2}) to (\ref{gl20}) backwards, it
becomes even more obvious that the $\eta$--coordinates in
(\ref{gl20}) describe gluons rather than mesons or
pseudo-mesons. In (\ref{neu2}) the $\eta$--coordinates are then
directly introduced as new names for the gluons.
Note that we do not need any kind of $\bar qq$--clustering, if
(\ref{gl20}) is derived in this way. In particular we would have
obtained equation (\ref{gl20}) without the $\eta$--quark coupling
term from the pure Yang--Mills action
$S\mscript{YM} = \int \dvier x \, \frac{1}{4} \, F^a_{\mu\nu}
F^a_{\mu\nu} $ instead of $S\mscript{QCD}$ in (\ref{gl1}).

\section{Conclusions}
The crucial point of our results is the triviality of the
bosonization procedure.

If path integral bosonization is taken
to be a meaningful tool for deriving effective
composite particle actions then, in QCD, we are {\it forced\/} to
regard the gluons as $\bar qq$--composites. This contradicts the
fundamentals of QCD, since gluons and quarks are introduced as
elementary and independent fields there, but by path integral
bosonization the gluons are turned into dependent composite
fields.

We must therefore reject path integral
bosonization as a technique to obtain serious composite particle
theories. The simple boson-like character of
coordinates (possibly associated with correct total quantum
numbers) in the path integral does not exhaust the rich structure
of a true composite particle theory in that case and cannot
guarantee for the derivation of mesons as observable states.

Further we have shown that ``mesons'' obtained by a Fierz
reordering and carrying different quantum numbers as the gluons
are restricted variables in the path integral. If the path
integral is evaluated correctly by bosonized Berezin rules, the
``mesons'' turn out to be gluons again. This shows that
colorlessness is not a sufficient condition for detecting mesons.

Note finally that no use was made of special properties of QCD.
Therefore our objections on path integral bosonization apply to
other models as well.

\end{document}